\documentclass[format=acmsmall, review=false, screen=true]{acmart}

\usepackage{multirow}
\usepackage{booktabs}
\usepackage[ruled]{algorithm2e}

\SetAlFnt{\small}
\SetAlCapFnt{\small}
\SetAlCapNameFnt{\small}
\SetAlCapHSkip{0pt}
\IncMargin{-\parindent}

\setcopyright{acmcopyright}
\acmJournal{PACMHCI}
\acmYear{2017}
\acmVolume{1}
\acmNumber{2}
\acmArticle{57}
\acmMonth{11}
\acmPrice{15.00}
\acmDOI{10.1145/3134692}

\received{April 2017}
\received[revised]{July 2017}
\received[accepted]{November 2017}

\def\mwd{\,\;\;}
\def\os{\:\:}
\def\ds{\os\os}
\def\ts{\mwd\mwd}

\begin{document}

\title[Service Providers of the Sharing Economy]
{Service Providers of the Sharing Economy: Who Joins and Who Benefits?}

\author{Qing Ke}
\affiliation{%
\institution{Indiana University, Bloomington}
}

\begin{abstract}
Many ``sharing economy'' platforms, such as Uber and Airbnb, have become
increasingly popular, providing consumers with more choices and suppliers a
chance to make profit. They, however, have also brought about emerging issues
regarding regulation, tax obligation, and impact on urban environment, and have
generated heated debates from various interest groups. Empirical studies
regarding these issues are limited, partly due to the unavailability of
relevant data. Here we aim to understand service providers of the sharing
economy, investigating who joins and who benefits, using the Airbnb market in
the United States as a case study. We link more than 211 thousand Airbnb
listings owned by 188 thousand hosts with demographic, socio-economic status
(SES), housing, and tourism characteristics. We show that income and education
are consistently the two most influential factors that are linked to the
joining of Airbnb, regardless of the form of participation or year. Areas with
lower median household income, or higher fraction of residents who have
Bachelor's and higher degrees, tend to have more hosts. However, when
considering the performance of listings, as measured by number of newly
received reviews, we find that income has a positive effect for entire-home
listings; listings located in areas with higher median household income tend to
have more new reviews. Our findings demonstrate empirically that the
disadvantage of SES-disadvantaged areas and the advantage of SES-advantaged
areas may be present in the sharing economy.
\end{abstract}




\thanks{Author's address: Q.~Ke, School of Informatics and Computing, Indiana
University, Bloomington, 919 East 10th Street, Bloomington, IN 47408, US}

\maketitle

\renewcommand{\shortauthors}{Q. Ke}

\section{Introduction}

Online peer-to-peer marketplaces, such as Uber, Airbnb, and TaskRabbit, connect
sellers with buyers through the exchange of goods or
services~\cite{Azevedo-matching-2016, Einav-p2p-2016, Luca-design-2016}. Widely
referred to as examples of the ``sharing economy''~\cite{Cusumano-sharing-2015,
Malhotra-sharing-2014}, they provide suppliers with new sources of income and
consumers diverse choices of products, and have become increasingly important
in the digital economy. They have also brought about emerging issues such as
regulation, tax obligation, and impact on neighborhood. Although there have
been intense debates from various interest groups on these issues, empirical
studies are limited, partly due to the unavailability of relevant data,
impeding timely evidence-based policy making.

In this work, we bring empirical data to bear on a key issue---service
providers---in the current debates around the sharing economy. Service
providers are, for example, driver-partners who provide ride-sharing services
on Uber, or hosts who provide accommodations on Airbnb, or ``taskers'' who
provide labors to do the errand on TaskRabbit. Recent years have seen an
exponential growth of the number of service providers in sharing economy
markets (e.g., Uber~\cite{Hall-labor-2016} and Airbnb~\cite{Ke-abb-2017}). Such
fast growth may be because participation in these markets has the potential to
generate new sources of income. Another reason may be that online markets offer
service providers a great flexibility, in the sense that they could decide when
to work, work for how long, and which task request to accept, etc. A recent
study showed that flexible work schedules make Uber drivers earn more than
twice compared to traditional work arrangements~\cite{Chen-value-2017}.

Understanding service providers in the sharing economy has interested
researchers across many disciplines. A great effort has been put into the
analysis of demographics, such as age, gender, and race~\cite{Hall-labor-2016,
Kooti-uber-2017}, as well as the demonstration of the existence of a gender
gap~\cite{Kricheli-Katze-cents-2016} and of gender and racial bias and
discrimination in online marketplaces~\cite{Edelman-digital-2014,
Hannak-bias-2017}. Despite these efforts, our understanding is still limited by
the lack of empirical and large-scale analysis on the social-economic status
aspect. This work contributes to the understanding of service providers of the
sharing economy by studying how social-economic status (SES) affects the
participation in and benefit from the sharing economy. In particular, we aim to
answer the following research questions:
\begin{enumerate}
\item \textbf{RQ1:} How are SES characteristics associated with service
providers of the sharing economy? Does such association vary across different
forms of sharing (e.g., entire home vs. private room)?
\item \textbf{RQ2:} How does the association change over time?
\item \textbf{RQ3:} How are SES characteristics associated with service
providers who actually benefit from the sharing economy?
\end{enumerate}
Here as a first step, we focus on Airbnb hosts in the United States as a case
study. This is based on two reasons. First, although Airbnb is a pioneering
example of the sharing economy, there have been few empirical studies about its
service providers. Second, currently US is Airbnb's largest
market~\cite{Ke-abb-2017}, and extensive SES and other characteristics data are
available, allowing us to approach our questions systematically.

Our research questions are important due to several theoretical and practical
reasons. First, participation in the sharing economy may be more appealing to
people with lower-income than to richer ones, as it may provide potential
revenue stream. Several studies have already pointed out that monetary
compensation is one of the main motivations for participation in the sharing
economy~\cite{Lampinen-hosting-2016, Ikkala-monetizing-2015,
Teodoro-motivations-2014}. Second, many sharing economy platforms greatly
reduce the costs associated with joining the markets. For instance, they have
provided a system to make the payment safe, have verified registered customers,
and have made it easy for would-be workers to join the platform. Therefore,
they may attract workers from across the income spectrum. Third, there are
still entry barriers to participate, such as having an underutilized room
located in a neighborhood where would-be guests are willing come, and having
the ability to deal with issues involved in managing listings. Therefore,
certain groups of individuals may not afford becoming a host. Fourth,
heterogeneous housing characteristics, such as location and decoration,
together with the existence of racial discrimination on Airbnb as demonstrated
recently~\cite{Edelman-digital-2014, Edelman-racial-2017}, may make some
listings more profitable, while others not, which may affect the joining of
Airbnb. On the practical side, Airbnb has repeatedly claimed that it has been
employed mainly by lower-income residents as their revenue supplement, which is
yet to be verified.

To answer our research questions, we use data crawled from Airbnb between May
and August, $2016$ and employ standard regression methods to control for other
characteristics, such as housing and attractiveness. Our modeling results show
that after controlling for demographic, housing, and attractiveness
characteristics, income has a strong negative effect on the participation in
Airbnb; areas with lower median household income tend to have more hosts. We
also find that education is another influential factor; areas where there are a
larger portion of residents with higher education degrees are associated with
more hosts. However, when we consider the performance of listings, as measured
by the number of newly received reviews, we find that income has a positive
effect on entire-home listings, even after controlling for listing- and
host-level characteristics. This means that entire-home listings located in
areas with higher income tend to have more new reviews. Our findings therefore
may suggest the disadvantage of SES-disadvantaged areas and the advantage of
SES-advantaged areas.

\section{Related Work}

In this section, we set the context for our work. First, we present existing
work analyzing service providers of the sharing economy. We then briefly
overview general analysis on sharing economy platforms.

\subsection{Analysis of Service Providers}

The question of why service providers join sharing economy platforms has been
one focus of several studies. An early work by Teodoro \emph{et al.} used
interviews to understand the motivations behind becoming workers on Gigwalk and
TaskRabbit~\cite{Teodoro-motivations-2014}. Later studies focused on other
markets, such as Uber~\cite{Lee-working-2015} and
Airbnb~\cite{Ikkala-monetizing-2015, Lampinen-hosting-2016}. All these studies
have pointed out that monetary compensation is one of the main motivations for
participation in the sharing economy. This finding plays a major role in
motivating our work presented here.

Another line of work has focused on potential issues and challenges associated
with being service providers. First, one requirement for service providers is to
establish trust, because it is a key prerequisite for many online
markets~\cite{Luca-design-2016}. Ma \emph{et al.} examined self-descriptions of
Airbnb hosts on their profiles and how they are related to perceived
trustworthiness~\cite{Ma-self-2017}. Second, certain groups of service
providers may be faced with discriminations. On Airbnb, black hosts charged
less than non-black hosts for the equivalent
rental~\cite{Edelman-digital-2014}. On eBay, women sellers received a smaller
number of bids and lower prices for the same
product~\cite{Kricheli-Katze-cents-2016}. On TaskRabbit, workers' gender and
race were correlated with their evaluations such as reviews and
ratings~\cite{Hannak-bias-2017}.

One closely related thread of research is recent work that started to ask the
question of who service providers of the sharing economy are. These works have
focused on the demographic perspective. Hall and Krueger, for example,
presented a comprehensive analysis of Uber's active driver-partners in the US
and found that their age and education are similar to the general
workforce~\cite{Hall-labor-2016}. Using email receipt data, Kooti \emph{et al.}
also examined demographics of Uber drivers~\cite{Kooti-uber-2017}. In contrast
to these work,  our work here only not focuses on Airbnb, a primary yet less
studied example of the sharing economy, but also emphasizes the SES aspect.

Regarding the SES perspective, although there are some
discussions~\cite{Dillahunt-sharing-2016} and participatory-design based
studies~\cite{Dillahunt-promise-2015} on how sharing economy platforms can
benefit low-income service providers, we still lack a quantitative and
large-scale empirical analysis.

The SES of geographic areas where the service will be performed is also a
factor. A study by Thebault-Spieker \emph{et al.} found that task requests from
the low-SES south side of the Chicago metropolitan area were less likely to be
accepted by TaskRabbit workers~\cite{Thebault-Spieker-avoiding-2015}.

\subsection{General Analysis of Sharing Economy Platforms}

A relatively dearth of work has considered analysis on different aspects of
online markets. Using email receipt data about Uber rides, Kooti \emph{et al.}
found that, among others, homophilous matches between riders and drivers
contributed to higher ratings of drivers and that an accurate prediction of
which riders or drivers will become active Uber users can be obtained using
early rides information~\cite{Kooti-uber-2017}. Chen \emph{et al.} investigated
Uber's surge pricing algorithm and reported the existence of short-lived spikes
of ``surge multiplier''~\cite{Chen-peeking-2015}.

Focusing on Airbnb, Ke presented a comprehensive statistical analysis of
listings on the entire Airbnb market and investigated factors related to
listings' success~\cite{Ke-abb-2017}. An important conclusion from that study
is the prevalence of commercial hosts who own multiple listings on
Airbnb~\cite{Ke-abb-2017}. This has motivated us to exclude those hosts from
our analysis. Quattrone \emph{et al.}, studied how geographic and demographic
characteristics contributed to the grow of Airbnb in
London~\cite{Quattrone-who-2016}.

\begin{figure}[t]
\centering
\includegraphics[width=0.49\textwidth]{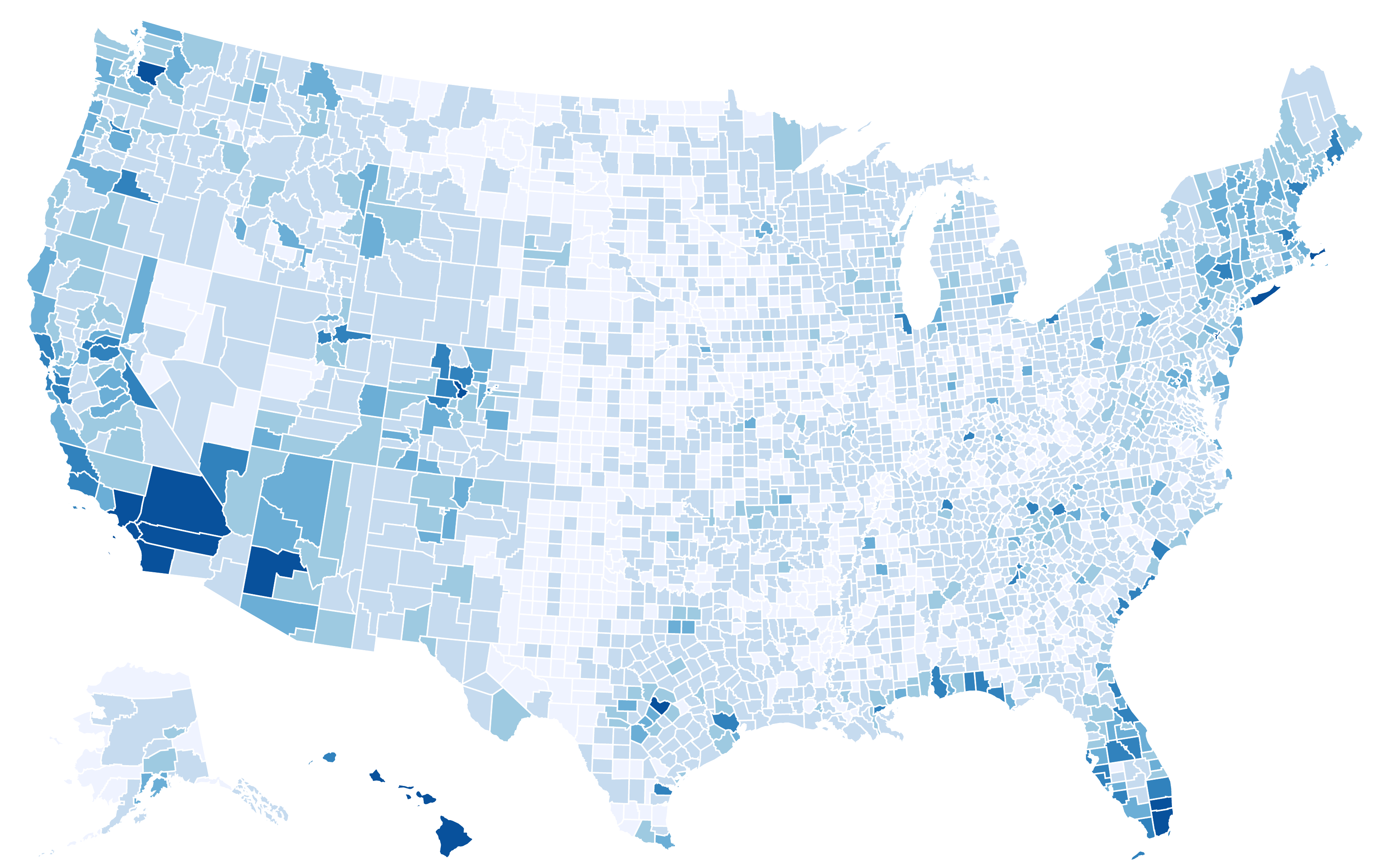}
\includegraphics[width=0.49\textwidth]{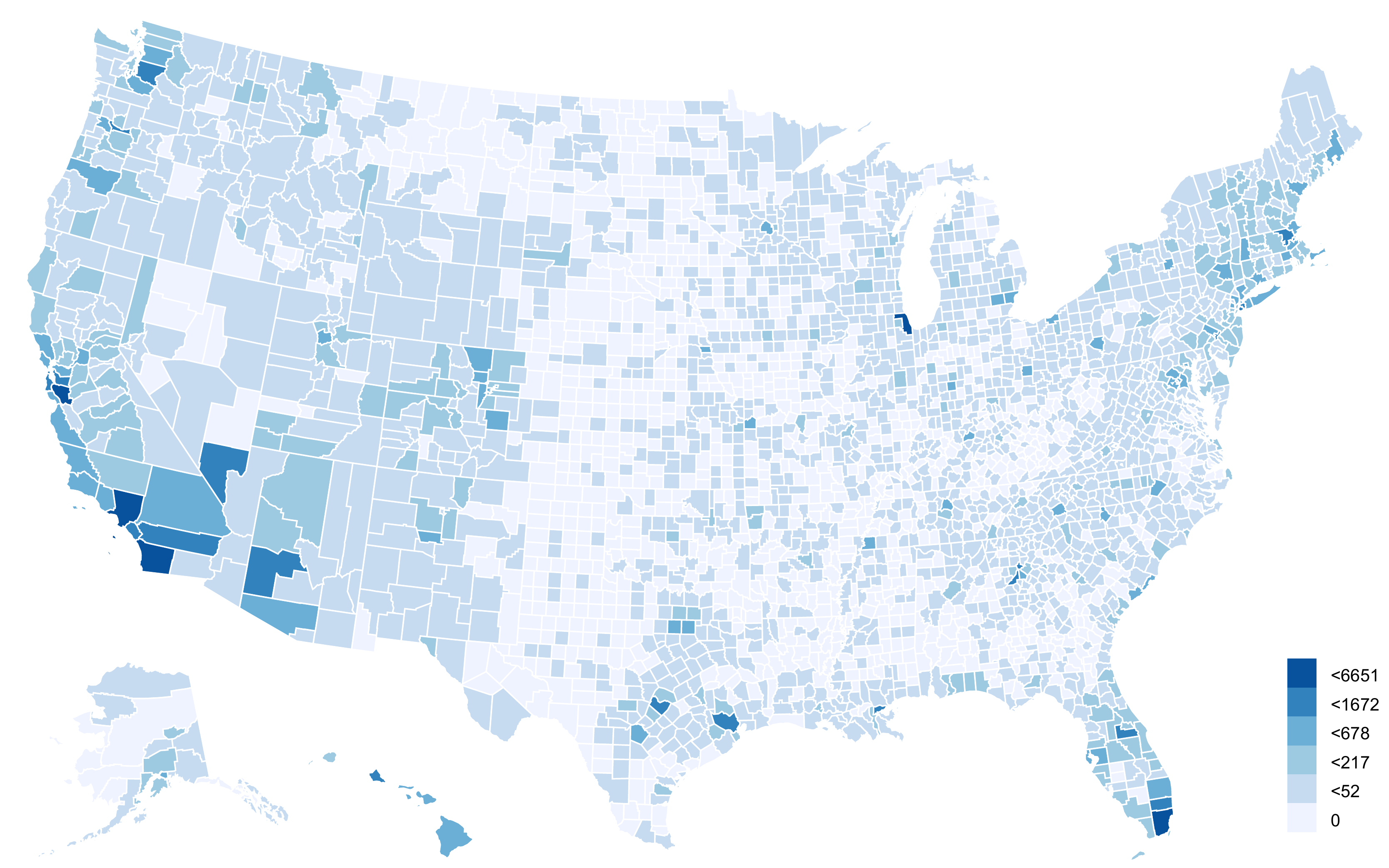}
\caption{Number of entire home listings (left) and private and shared room
listings (right) by county.}
\label{fig:loc}
\end{figure}

\section{The Airbnb Platform}

Our empirical study is set in the context of Airbnb, an online marketplace for
accommodation. Here we briefly describe how Airbnb works. Founded in $2008$,
Airbnb allows hosts to rent their underutilized rooms to guests. To become a
host, one can sign up as a host and then set up their listing by filling out
description, uploading photos, and setting a price and the calendar. A listing
can be one of three types of rooms, namely entire home, private room, and
shared room. Guests can find out listings mainly through Airbnb's search engine
and request to book the listing they intend to stay. After the conclusion of
stays, both hosts and guests can review and rate each other. However, reviews
and ratings are simultaneously revealed to avoid retaliation.

\section{Data and Methods}

\subsection{Airbnb Data}

We obtain the Airbnb data directly from a previous work~\cite{Ke-abb-2017}.
Here we briefly describe the data collection process, and refer to the original
study for more details~\cite{Ke-abb-2017}. Our data was crawled from the Airbnb
website. At the time of the crawl, Airbnb provided site maps with links to web
pages of search results of listings in all regions within countries that were
covered by the service. The crawling process first repeatedly queried those
search result pages to accumulate listings and then visited the web page of
each listing to obtain relevant information (e.g., host, ratings, and reviews).
A large fraction of listings was visited for three times during May and August,
$2016$, therefore we know how certain information such as ratings and
number of reviews change over time. This allows us to study the performances of
listings presented in \S~\ref{subsec:rq3}.

Overall, the data collection process resulted in more than $2.3$ million
listings---a number comparable to the official one as of the time data were
collected---on the entire Airbnb platform. Among them, $312,358$ ($13.6\%$) are
located in the US, making it the largest Airbnb market in the world.

To explore geolocations of listings, Fig.~\ref{fig:loc} displays the number of
entire home listings (left) and private and shared room listings (right) in
each county. Listings are heavily located along the east and west coast, and
there are many mid-west counties without any listings.

\begin{table}[t]
\caption{Statistics of Airbnb in the US.}
\label{tab:stats}
\begin{minipage}{\columnwidth}
\begin{center}
\begin{tabular}{l r r}
\toprule
         & Total     & This analysis \\
\midrule
Listings & $312,358$ & $211,124$ ($67.6\%$) \\
Hosts    & $203,612$ & $188,215$ ($92.4\%$) \\
\bottomrule
\end{tabular}
\end{center}
\end{minipage}
\end{table}

\subsubsection{Exclusion of Commercial Hosts}

A key feature of many online peer-to-peer markets is that on their supply side,
there are not only ordinary players who, in the context of Airbnb, rent out
their apartments, but also commercial hosts who operate business on the
platform. Commercial hosts, also referred to as professional hosts, are often
identified as those who own multiple listings on Airbnb. Currently they are
amid the focuses of the ongoing debates and one of the main targets of
regulations in many cities. They are, however, not the main subjects of our
study, as we are interested in how SES of \emph{local residents} affect their
joining in Airbnb. We therefore need to exclude commercial hosts from our
analysis.

To demonstrate the existence of commercial hosts, we show in
Fig.~\ref{fig:l-char}(a) the survival distribution of the number of listings
owned by a host and the number of unique census tracts (CT, a geographic
region, \emph{cf.} \S~\ref{subsubsec:dep-rq1}) where these listings are
located. We clearly see heavy-tailed distributions; among the $203,612$ hosts
in total, $165,306$ ($81.2\%$) and $22,909$ ($11.2\%$) of them own one and two
listings, respectively, the remaining $7.6\%$ hosts own $32.4\%$ listings, and
one host even has $855$ listings. The number of CTs where a host's listings are
located can reach $424$. These results, consistent with previous observations
focusing on the global scale~\cite{Ke-abb-2017}, indicate the existence of
commercial hosts.

Based on this observation, we exclude hosts who have more than two listings and
exclude their listings before further analysis. In this way, we are still able
to include a majority ($92.4\%$) of hosts who are most likely to be ordinary
residents rather than commercial operators. After the exclusion, there are only
$15,949$ ($8.5\%$) hosts whose listings are located in two different CTs.
Table~\ref{tab:stats} summaries basic statistics of our dataset.

\subsubsection{Listing Characteristics}

\begin{figure}[t]
\centering
\includegraphics[width=0.7\columnwidth]{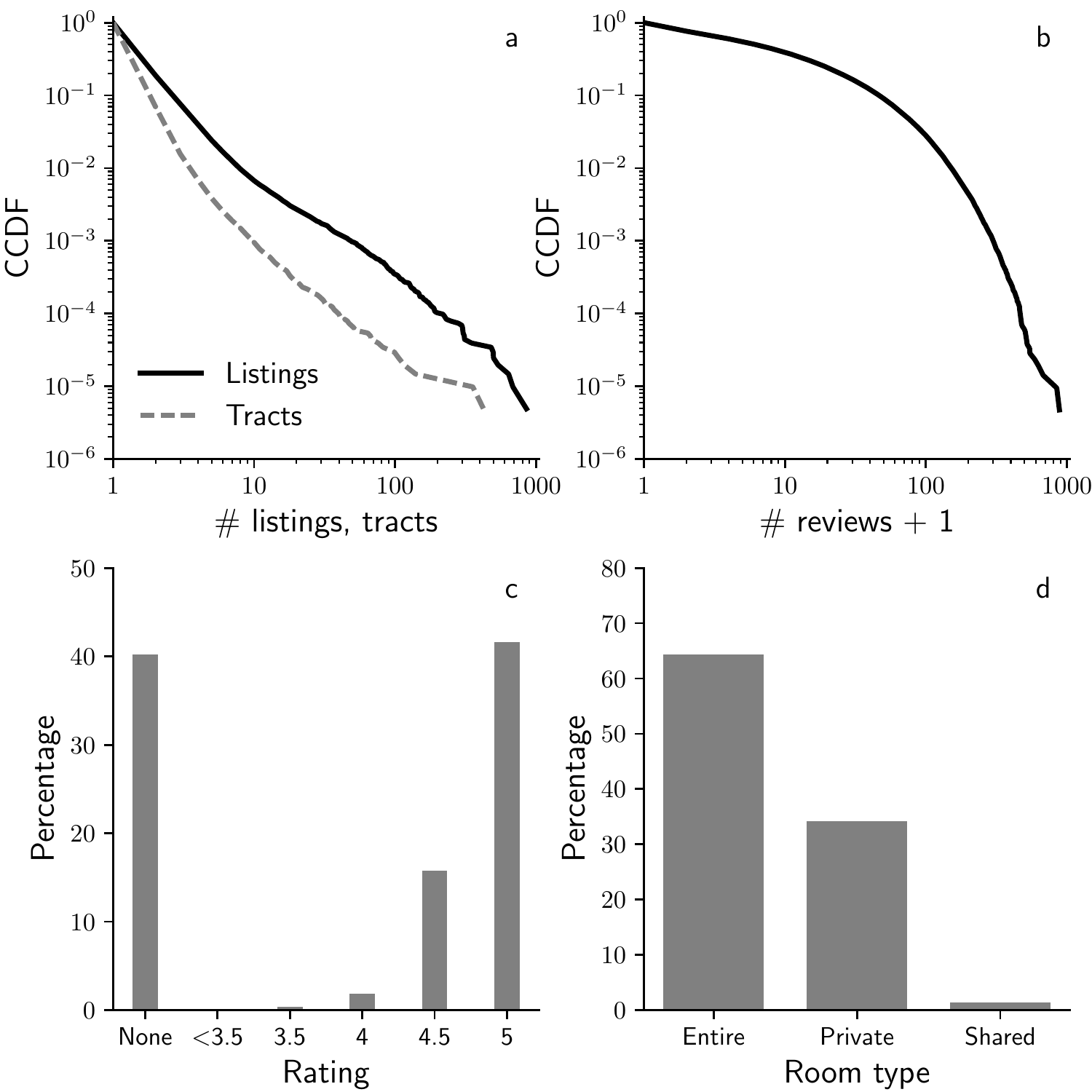}
\caption{(a) Survival distribution of the number of listings owned by a host
and the number of census tracts where these listings are located. We exclude
hosts with more than two listings before further analysis. After the exclusion,
we show the distribution of number of reviews (b), rating (c), and room type
(d).}
\label{fig:l-char}
\end{figure}

We then present a basic summary of listing characteristics.
Figure~\ref{fig:l-char}(b) shows the heterogeneously distributed number of
reviews across listings. The mean (median) number of reviews received by a
listing is $16.2$ ($5$). Figure~\ref{fig:l-char}(c) demonstrates a bimodal
distribution of ratings. There are roughly the same number of no-rating
listings and five-star listings, while listings with low ratings rarely exist
on the market. Figure~\ref{fig:l-char}(d) shows that more than $60\%$ of
listings are entire homes.

\subsection{Dependent Variables}

We now introduce the dependent variables and show their distributions.

\subsubsection{RQ1} \label{subsubsec:dep-rq1}

\begin{figure}[t]
\centering
\includegraphics[width=0.7\columnwidth]{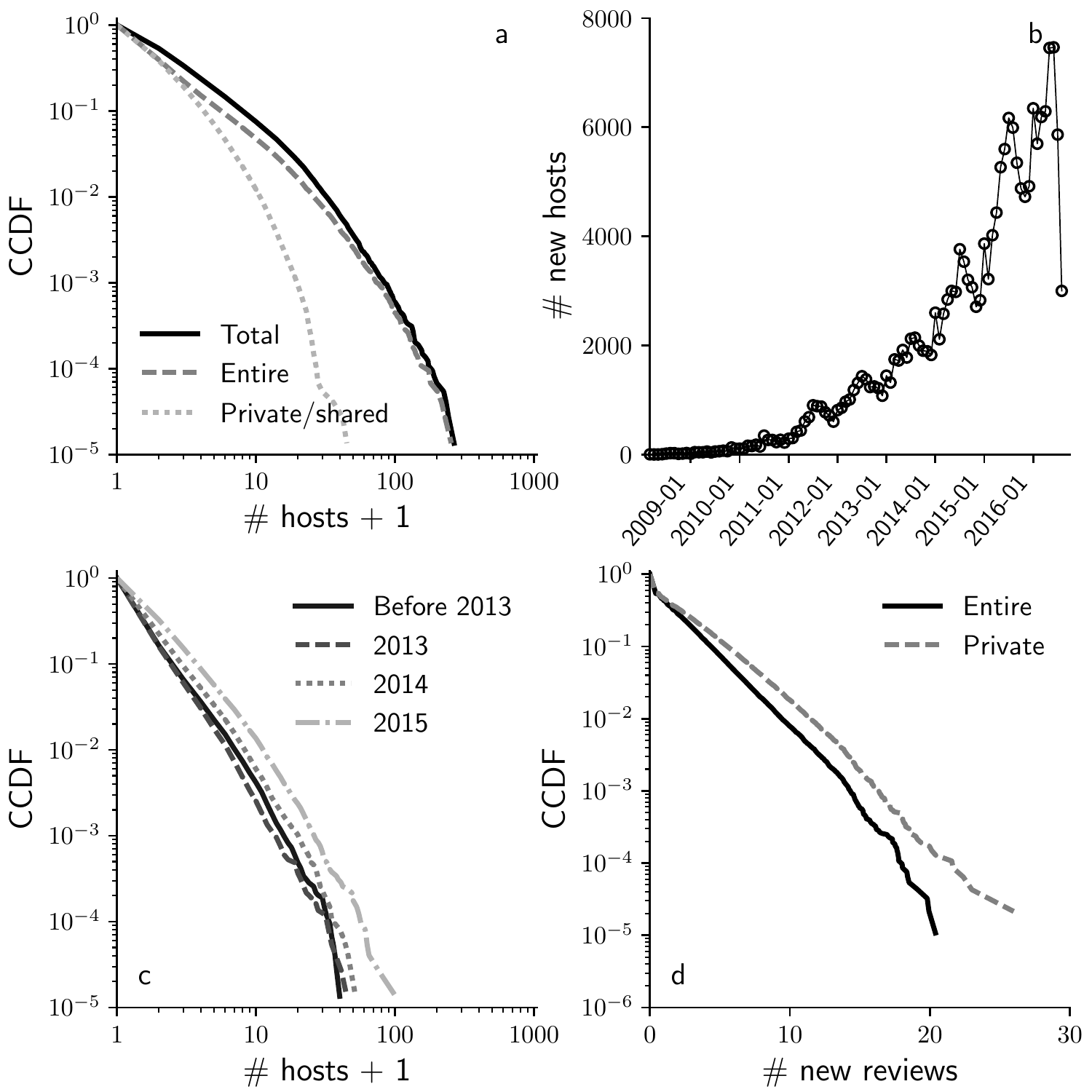}
\caption{(a) Distribution of the number of unique hosts in a census tract; (b)
Number of new hosts over time; (c) Distribution of the number of unique hosts
in a census tract by joining time; (d) Distribution of monthly new reviews of
entire homes and private rooms.}
\label{fig:depen-dist}
\end{figure}

RQ1 asks how SES characteristics are associated with service providers who
provide accommodations on the Airbnb platform. The ideal unit of analysis is
individual. This is, however, unrealistic, because doing so would not only
require detailed individual SES data that are difficult to obtain but also
raise privacy concerns.

We choose to perform our analysis at the census tract (CT)
level.\footnote{\url{https://www.census.gov/geo/reference/gtc/gtc_ct.html}}
CT is a geographic region within a county and specifically designed to present
statistical data such as median household income that shall be used extensively
in our study. The primary advantage of CT is that it is a small, relatively
homogeneous, and stable geographic area, which allows for statistical
comparisons between different censuses. By focusing on the CT level, we are
making two assumptions: (1) hosts are local residents of the CT where their
listings are located; and (2) the SES characteristics of a CT is representative
of its residents.

The dependent variable for RQ1 is then the number of unique hosts in a CT. To
obtain the number, we map each US listing to the CT it is located, using its
latitude and longitude information and the Geocoder
API\footnote{\url{https://geocoding.geo.census.gov/geocoder/}} provided by the
US Census Bureau.

Since Airbnb distinguishes between different types of rooms, we also
investigate variations among rooms types by looking at the number of hosts of
entire rooms and of private and shared rooms. In summary, we use three
dependent variables for RQ1: (1) number of unique hosts, (2) number of unique
hosts of entire home listings, and (3) number of unique hosts of private and
shared rooms.

Figure~\ref{fig:depen-dist}(a), which shows the survival distributions of the
three variables, suggests that Airbnb has heterogeneous popularities among CTs.
The mean (median) number of hosts in a CT is $2.6$ ($1$), and the maximum
number is less than $300$. By contrast, the average total population in a CT is
$4,329$, suggesting that the overall adoption rate of Airbnb is not high.

\subsubsection{RQ2}

\begin{table}[t]
\caption{List of independent variables, their definition, data source, and
basic statistics.}
\label{tab:var}
\begin{minipage}{\columnwidth}
\begin{center}
\begin{tabular}{l l l r r}
\toprule
Variable & Definition & Source & Mean & Median \\
\midrule
$popu$ & Total population & S0101 & $4,329$ & $4,067$ \\
$young$ & Fraction of residents aged between $20$--$34$ & S0101 & $0.207$ & $0.193$ \\
$nonwhite$ & Fraction of non-white residents & B02001 & $0.269$ & $0.184$ \\
$foreign$ & Fraction of foreign-born residents & B05003 & $0.151$ & $0.084$ \\
\midrule
$income$ & Median household income & S1903 & $57,049$ & $50,972$ \\
\multirow{2}{*}{$edu$} & Fraction of 25+ years old residents with & \multirow{2}{*}{B15003} & \multirow{2}{*}{$0.285$} & \multirow{2}{*}{$0.234$} \\
& \hspace{5mm} Bachelor's and higher degree & & & \\
$employment$ & Employment rate & B23025 & $0.910$ & $0.923$ \\
\multirow{2}{*}{$arts$} & Fraction of employed residents working in arts, & \multirow{2}{*}{S2404} & \multirow{2}{*}{$0.069$} & \multirow{2}{*}{$0.054$} \\
& \hspace{5mm} entertainment, etc & & & \\
\midrule
$owned$ & Fraction of owner-occupied housing units & B25003 & $0.632$ & $0.683$ \\
$housing\_value$ & Median price of housing units occupied by owner & B25077 & $222,771$ & $162,400$ \\
$rent$ & Median rent of housing units occupied by renter & B25064 & $996.6$ & $890.0$ \\
\midrule
$hotel$ & Number of hotels & OSM & $0.231$ & $0$ \\
\multirow{2}{*}{$attraction$} & Number of attractions, retails, bars, and & \multirow{2}{*}{OSM} & \multirow{2}{*}{$1.199$} & \multirow{2}{*}{$0$} \\
& \hspace{5mm} restaurants, etc & & & \\
\bottomrule
\end{tabular}
\end{center}
\bigskip
\footnotesize\emph{Note:} For the source column, OSM means OpenStreetMap and
for all others, we show the Table IDs used to retrieve corresponding data from
the American FactFinder website.
\end{minipage}
\end{table}

RQ2 asks how the impact of SES factors on participation in the sharing economy
changes over time. To introduce dependent variables, we first show in
Fig.~\ref{fig:depen-dist}(b) the number of newly joined hosts in each month
from March, $2008$ to August, $2016$. We see that Airbnb has experienced rapid
growth since around year $2012$. As we only have partial data for $2016$, we
exclude them when answering RQ2 and divide the entire $2008$--$2015$ period
into four periods, namely, before $2013$, $2013$, $2014$, and $2015$. The
number of hosts for each period are as follows: before $2013$: $24,460$;
$2013$: $21,798$; $2014$: $35,210$; and $2015$: $58,437$. The dependent
variables are the number of newly joined hosts in each period.
Figure~\ref{fig:depen-dist}(c) shows the distribution of the number of unique
hosts in a CT for each period.

For both RQ1 and RQ2, we use negative binomial model, since the number of
unique hosts is a count variable.

\subsubsection{RQ3}

So far, we have only considered who participate in the sharing economy. However,
not all listings are equally profitable, corroborated by the heterogeneous
distribution of the number of reviews across listings
(Fig.~\ref{fig:l-char}(b)). RQ3 asks who actually benefit from the sharing
economy. A straightforward measure of such benefit is the amount of revenues
generated by renting out listings. The calculation of revenues, however,
requires not only the proprietary listing occupancy data, but also historical
price information that are hard to obtain. Another measure could be the number
of reviews a listing has already received. This is also related to many
time-evolving factors that we have not observed, e.g., when a host became a
\texttt{superhost}, a special type of host who satisfies a series of
requirements.

To answer RQ3, we take advantage of the fact that the dataset contains three
snapshots of Airbnb listings. We then measure the performance as the number of
new reviews received between the first and third snapshots, and use it as the
dependent variable. As the number of days between the two snapshots varies
across listings, we further normalize performance into the number of new
reviews in a month, treating it as a continuous variable. We then regress
performance on the independent variables measured at the first snapshot using
ordinary least squares (OLS) regression.

Figure~\ref{fig:depen-dist}(d) shows the distribution of monthly new reviews of
entire home listings and private room listings.

\subsection{Independent Variables}

We now describe the list of independent variables used in our study. To
understand how SES factors are associated with the participation in the sharing
economy, we need to control for other factors such as demographics, housing,
and attractiveness. We rely on two sources---American Community Survey (ACS)
and OpenStreetMap (OSM)---to obtain these independent variables.
Table~\ref{tab:var} lists all these variables used in our analysis and their
basic statistics. Below we describe them in detail.

\subsubsection{Demographic, SES, and Housing Variables}

We collect demographic, SES, and housing variables about CTs from the
$2011$--$2015$ $5$-Year Estimates of American Community Survey (ACS). These
data are directly accessible through the American FactFinder website at
\url{https://factfinder.census.gov} by providing Table IDs. To improve
reproducibility, the ``Source'' column in Table~\ref{tab:var} lists the Table
IDs that were used to retrieve these variables from the website.

We include four variables characterizing the demographic features of a CT.
They are: (1) total population ($popu$), which accounts for the effect that
populous CTs tend to have more hosts by pure chance, (2) fraction of residents
whose age is between $20$ and $34$ ($young$), (3) fraction of non-white
residents ($nonwhite$), which is included based on previous results showing
discrimination in online markets~\cite{Edelman-digital-2014, Hannak-bias-2017,
Kricheli-Katze-cents-2016}, and (4) fraction of foreign-born residents
($foreign$).

\begin{figure}[t]
\centering
\includegraphics[width=\textwidth]{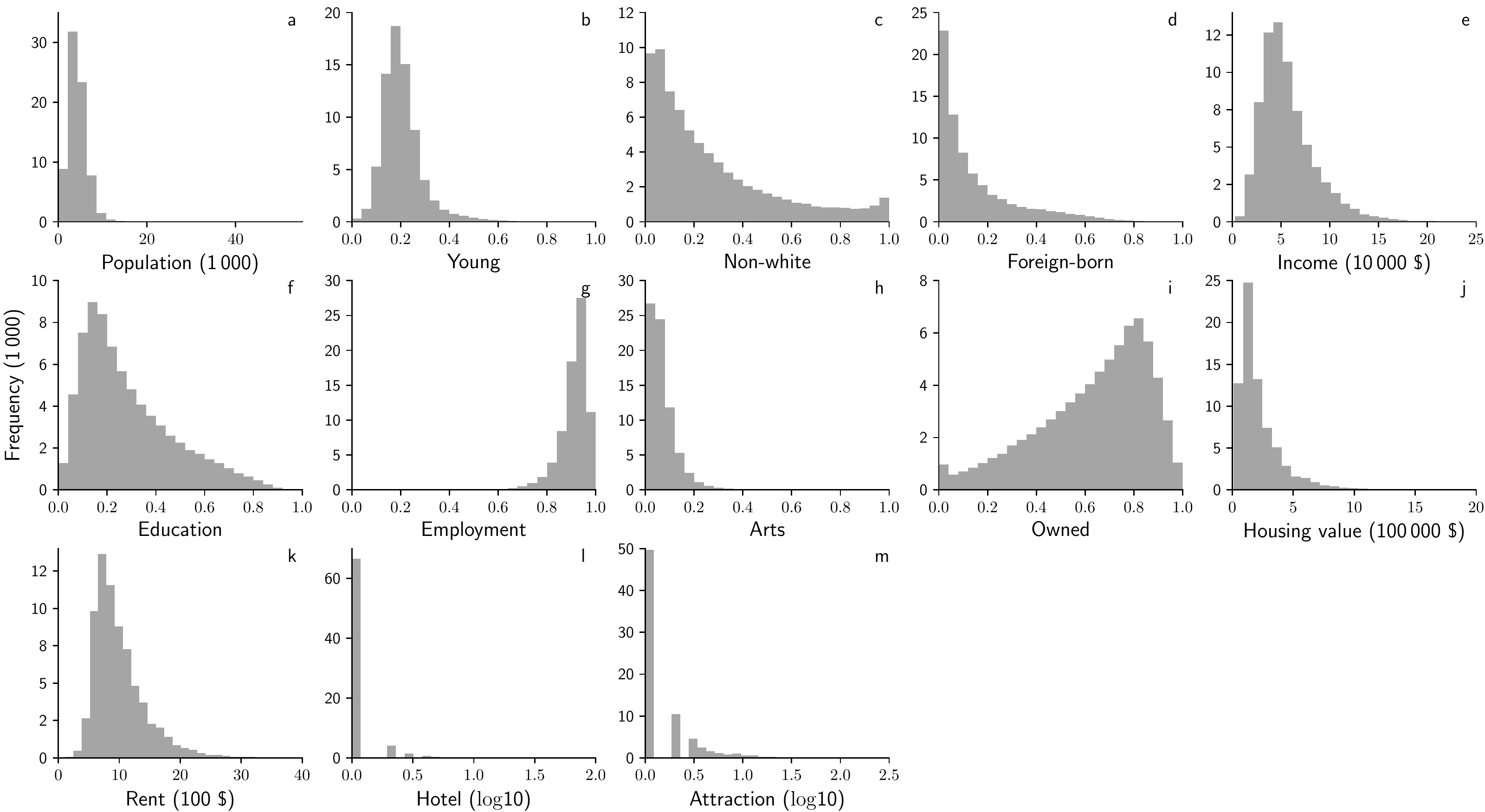}
\caption{Distribution of independent variables about demographic,
socio-economic status, housing, and tourism characteristics.}
\label{fig:indepen-dist}
\end{figure}

SES measures are often based on income, education, and occupation. We include
four variables into our analysis: (1) median household income ($income$), which
has been shown to be one factor in joining various
platforms~\cite{Teodoro-motivations-2014, Lee-working-2015,
Ikkala-monetizing-2015, Lampinen-hosting-2016}, (2) fraction of 25+ years old
residents with Bachelor's or higher degrees ($edu$), (3) employment rate
($employment$), and (4) fraction of employed residents working in arts,
entertainment, and other tourism-related industries ($arts$).

We include three variables characterizing housing conditions in a CT. They are
(1) fraction of owner-occupied housing units ($owned$), (2) median price of
housing units occupied by owner ($housing\_value$), and (3) median rent of
housing units occupied by renter ($rent$).

\begin{figure}[t]
\centering
\includegraphics[trim=0mm 10mm 0mm 10mm, clip, width=0.6\columnwidth]{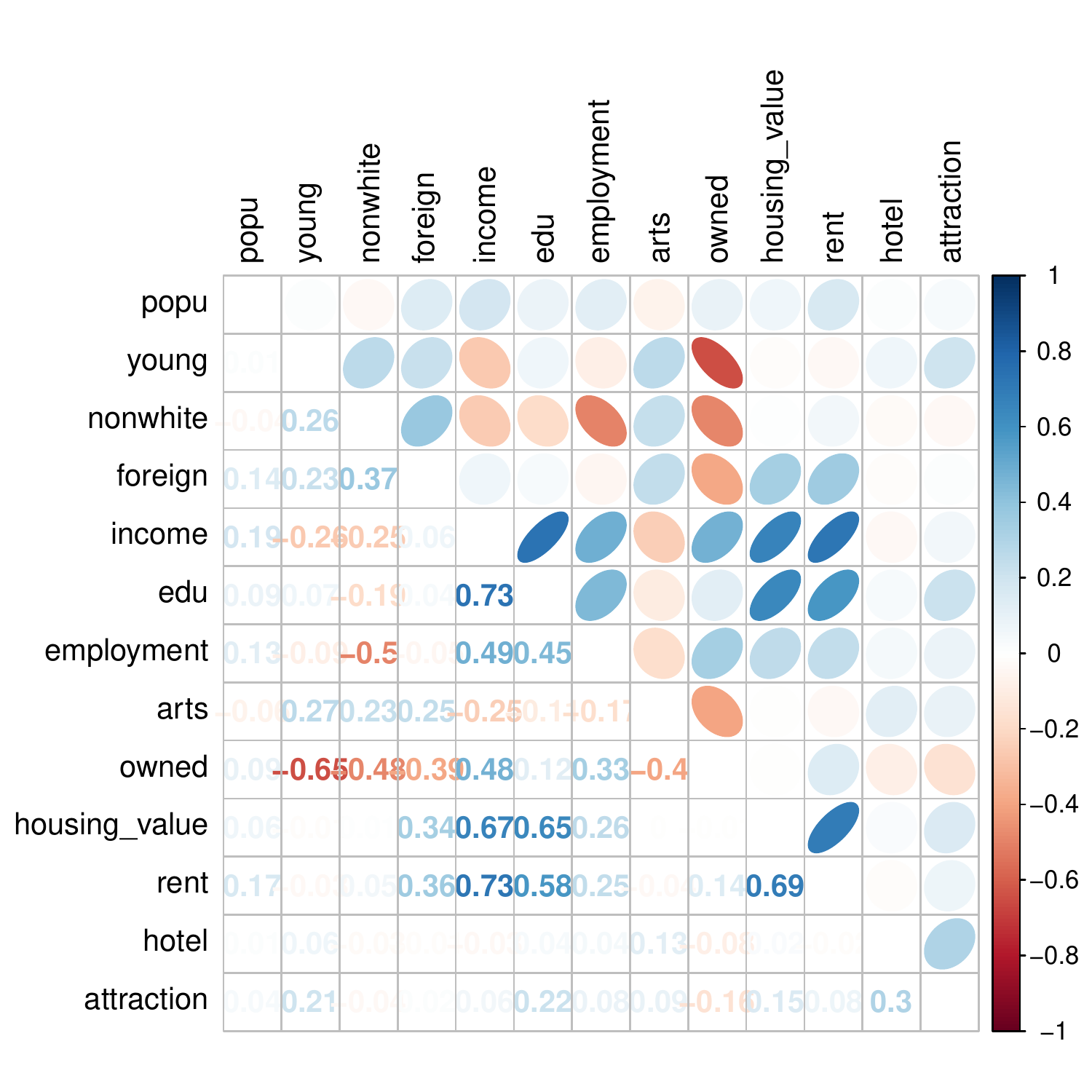}
\caption{Correlation matrix of independent variables.}
\label{fig:corr-mtrx}
\end{figure}

Figures~\ref{fig:indepen-dist}(a)--(k) show the distributions of these
variables. As we can see, they are not extremely heavily-tailed, since they are
statistics of summary statistics in a CT.

\subsubsection{Tourism Variables}

As the attractiveness of a CT may also impact its residents' tendency to join
Airbnb, we need to control for such attractiveness. Furthermore, Airbnb has
claimed that listings are located outside main hotel areas. To test the claim
and to study how Airbnb listing locations are different from hotel locations,
we collect hotel and attraction data from OpenStreetMap (OSM). In particular,
we use OSM Overpass
API\footnote{\url{http://wiki.openstreetmap.org/wiki/Overpass_API}} and make
queries, for each state, using its bounding box and the querying tags. For
hotel data, we use the following tags: ``tourism=hotel'', ``tourism=hostel'',
``tourism=motel'', and ``building=hotel''; and for attraction data:
``tourism=aquarium'', ``tourism=artwork'', ``tourism=attraction'',
``tourism=museum'', ``tourism=zoo'', ``building=retail'', ``amenity=bar'', and
``amenity=restaurant''. We then map the set of returned hotels and attractions
to CTs using the Geocoder API, and count the number of hotels and attractions
in each CT, denoted as $hotel$ and $attraction$, respectively.
Figures~\ref{fig:indepen-dist}(l)--(m) shows the distributions of the two
variables. Most CTs do not have any hotels or attractions.

Figure~\ref{fig:corr-mtrx} reports correlations between these independent
variables. As expected, some of them are correlated, such as rent and income,
and rent and housing value. For ease of interpretation and comparison across
them, we standardize all independent variables.

\section{Results}

\subsection{RQ1}

\setlength{\tabcolsep}{14pt}
\begin{table}[t]
\caption{Negative binomial regressions of number of hosts in a census tract.}
\label{tab:nhost-regress}
\begin{minipage}{\columnwidth}
\begin{center}
\begin{tabular}{l c c c}
\toprule
& \multicolumn{3}{c}{\emph{Dependent variable: Number of hosts}} \\
\cline{2-4}
                 & All                       & Entire home               & Private/shared room         \\
\midrule
$popu$           & $\mwd0.220^{***}$ (0.006) & $\mwd0.173^{***}$ (0.007) & $\mwd0.293^{***}$   (0.006) \\
$hotel$          & $\mwd0.122^{***}$ (0.005) & $\mwd0.144^{***}$ (0.006) & $\mwd0.055^{***}$   (0.005) \\
$attraction$     & $\mwd0.137^{***}$ (0.006) & $\mwd0.170^{***}$ (0.007) & $\mwd0.061^{***}$   (0.005) \\
$young$          & $   -0.061^{***}$ (0.008) & $   -0.102^{***}$ (0.010) & $\mwd0.047^{***}$   (0.008) \\
$nonwhite$       & $   -0.165^{***}$ (0.008) & $   -0.285^{***}$ (0.010) & $\mwd0.024^{**\os}$ (0.008) \\
$foreign$        & $   -0.095^{***}$ (0.008) & $   -0.192^{***}$ (0.010) & $\mwd0.017^{*\ds}$  (0.008) \\
$owned$          & $\mwd0.139^{***}$ (0.011) & $\mwd0.200^{***}$ (0.014) & $   -0.022^{\ts}$   (0.012) \\
$housing\_value$ & $\mwd0.553^{***}$ (0.009) & $\mwd0.666^{***}$ (0.011) & $\mwd0.293^{***}$   (0.008) \\
$rent$           & $\mwd0.216^{***}$ (0.010) & $\mwd0.171^{***}$ (0.012) & $\mwd0.240^{***}$   (0.010) \\
$income$         & $   -0.915^{***}$ (0.014) & $   -1.056^{***}$ (0.018) & $   -0.570^{***}$   (0.014) \\
$edu$            & $\mwd0.870^{***}$ (0.010) & $\mwd0.911^{***}$ (0.012) & $\mwd0.741^{***}$   (0.010) \\
$employment$     & $\mwd0.009^{\ts}$ (0.008) & $   -0.019^{\ts}$ (0.010) & $\mwd0.065^{***}$   (0.009) \\
$arts$           & $\mwd0.367^{***}$ (0.006) & $\mwd0.418^{***}$ (0.008) & $\mwd0.258^{***}$   (0.006) \\
Constant         & $\mwd0.460^{***}$ (0.006) & $   -0.042^{***}$ (0.007) & $   -0.501^{***}$   (0.007) \\
\midrule
Observations     & $69,824$ & $69,824$ & $69,824$ \\
Log Likelihood   & $-123,799.6$ & $-97,755.8$ & $-79,350.3$ \\
$\theta$         & $0.637^{***}$ ($0.005$) & $0.431^{***}$ ($0.004$) & $1.023^{***}$ ($0.014$) \\
AIC              & $247,627.2$ & $195,539.6$ & $158,728.6$ \\
\bottomrule
\end{tabular}
\end{center}
\bigskip
\footnotesize\emph{Note:} \hfill $^{*}\text{p}<0.05$; $^{**}\text{p}<0.01$; $^{***}\text{p}<0.001$
\end{minipage}
\end{table}

The ``All'' column in Table~\ref{tab:nhost-regress} reports the results from
negative binomial regression where the dependent variable is the number of
unique hosts in a CT and the independent variables characterize demographic,
SES, housing, and tourism features of that CT. Unsurprisingly, total population
is positively associated with the number of hosts, and so is the attractiveness
of CTs measured by number of hotels and attractions. CTs with lower fraction of
young, non-white, or foreign-born residents tend to have more hosts. In terms
of housing conditions, CTs with higher fraction of housing units occupied by
owners, higher median housing price, or higher rent on average have larger
number of hosts. Finally, after controlling for all these variables, income is
negatively correlated with number of hosts, indicating that, all else equal,
CTs with lower median household income tend to have more residents who have
joined Airbnb as hosts. Note that income is the most influential predictor; CTs
with median household income that is one standard deviation below the average
have $0.915$ more host. The overall education level of a CT has the second
largest effect on joining Airbnb; areas where there are a larger portion of
residents with higher education degree have more hosts.

We further investigate whether the association patterns discussed above vary
across room types. To do so, we repeat the same procedure above but separate
hosts of entire homes and hosts of private and shared rooms. The last two
columns in Table~\ref{tab:nhost-regress} display the results. Most association
patterns persist. Income and education remain to have the largest effects.
Housing price, rent, and fraction of residents working in arts and other
related industries are still positively correlated with number of hosts
regardless of room type. Demographics factors exhibit opposite patterns for
entire home hosts and private/shared room hosts; CTs with higher fraction of
young, nonwhite, or foreign-born residents have more hosts of private and
shared rooms, but less hosts of entire homes.

\subsection{RQ2}

Next, we discuss how the association patterns change over time.
Table~\ref{tab:nhost-entire-year} and \ref{tab:nhost-other-year} report the
results from negative binomial regressions where dependent variables are,
respectively, the number of entire home hosts and private/shared room hosts in
a CT who joined Airbnb in different periods, and the independent variables are
demographic, SES, housing, and tourism features of that CT.

Table~\ref{tab:nhost-entire-year} reports the results for entire home hosts.
First, we note that the directions of association for all variables are the
same as the case where we lump together all entire home hosts joining Airbnb in
different years (the ``Entire home'' column in Table~\ref{tab:nhost-regress}),
and are consistent across years. Income and education remain to be the two most
influential factors across years. All else equal, CTs with lower median
household income or higher fraction of residents with Bachelor+ degrees tend to
have more hosts. Housing value has the third largest effect and is positively
linked to the number of hosts. Such pattern is consistent across years,
although the coefficient steadily decreases over time. The coefficient for the
rent variable keeps rising, which may suggest that Airbnb is increasingly
adopted in areas with higher rents.

\setlength{\tabcolsep}{6pt}
\begin{table}[t]
\caption{Negative binomial regressions of number of entire home hosts in a
census tract over time.}
\label{tab:nhost-entire-year}
\begin{minipage}{\columnwidth}
\begin{center}
\begin{tabular}{l c c c c}
\toprule
& \multicolumn{4}{c}{\emph{Dependent variable: Number of hosts (entire home)}} \\
\cline{2-5}
                 & Before 2013               & 2013                      & 2014                      & 2015 \\
\midrule
$popu$           & $\mwd0.110^{***}$ (0.013) & $\mwd0.117^{***}$ (0.013) & $\mwd0.131^{***}$ (0.011) & $\mwd0.167^{***}$ (0.009) \\
$hotel$          & $\mwd0.104^{***}$ (0.009) & $\mwd0.109^{***}$ (0.009) & $\mwd0.126^{***}$ (0.008) & $\mwd0.134^{***}$ (0.007) \\
$attraction$     & $\mwd0.148^{***}$ (0.009) & $\mwd0.135^{***}$ (0.009) & $\mwd0.143^{***}$ (0.008) & $\mwd0.147^{***}$ (0.007) \\
$young$          & $   -0.117^{***}$ (0.016) & $   -0.113^{***}$ (0.016) & $   -0.090^{***}$ (0.014) & $   -0.083^{***}$ (0.012) \\
$nonwhite$       & $   -0.281^{***}$ (0.019) & $   -0.293^{***}$ (0.019) & $   -0.323^{***}$ (0.016) & $   -0.273^{***}$ (0.013) \\
$foreign$        & $   -0.172^{***}$ (0.018) & $   -0.186^{***}$ (0.018) & $   -0.205^{***}$ (0.015) & $   -0.222^{***}$ (0.013) \\
$owned$          & $\mwd0.169^{***}$ (0.024) & $\mwd0.193^{***}$ (0.024) & $\mwd0.208^{***}$ (0.021) & $\mwd0.189^{***}$ (0.018) \\
$housing\_value$ & $\mwd0.792^{***}$ (0.016) & $\mwd0.727^{***}$ (0.016) & $\mwd0.677^{***}$ (0.015) & $\mwd0.526^{***}$ (0.013) \\
$rent$           & $\mwd0.113^{***}$ (0.020) & $\mwd0.157^{***}$ (0.020) & $\mwd0.186^{***}$ (0.017) & $\mwd0.174^{***}$ (0.015) \\
$income$         & $   -1.100^{***}$ (0.029) & $   -1.110^{***}$ (0.029) & $   -1.093^{***}$ (0.026) & $   -0.981^{***}$ (0.022) \\
$edu$            & $\mwd0.929^{***}$ (0.020) & $\mwd0.899^{***}$ (0.020) & $\mwd0.878^{***}$ (0.018) & $\mwd0.879^{***}$ (0.015) \\
$employment$     & $ -0.059^{**\os}$ (0.018) & $   -0.024^{\ts}$ (0.018) & $   -0.005^{\ts}$ (0.016) & $   -0.004^{\ts}$ (0.013) \\
$arts$           & $\mwd0.442^{***}$ (0.012) & $\mwd0.414^{***}$ (0.012) & $\mwd0.433^{***}$ (0.011) & $\mwd0.421^{***}$ (0.009) \\
Constant         & $   -2.171^{***}$ (0.015) & $   -2.234^{***}$ (0.015) & $   -1.728^{***}$ (0.012) & $   -1.151^{***}$ (0.010) \\
\midrule
Observations     & $69,824$ & $69,824$ & $69,824$ & $69,824$ \\
Log Likelihood   & $-31,308.080$ & $-29,930.890$ & $-40,527.930$ & $-55,844.330$ \\
$\theta$         & $0.306^{***}$ (0.007) & $0.343^{***}$ (0.008) & $0.352^{***}$ (0.007) & $0.413^{***}$ (0.006) \\
AIC              & $62,644.150$ & $59,889.780$ & $81,083.850$ & $111,716.700$ \\
\bottomrule
\end{tabular}
\end{center}
\bigskip
\footnotesize\emph{Note:} \hfill $^{*}\text{p}<0.05$; $^{**}\text{p}<0.01$; $^{***}\text{p}<0.001$
\end{minipage}
\end{table}

Table~\ref{tab:nhost-other-year} provides the regression results for
private/shared room hosts. Consistent with all previous results, income and
education are the two most important predictors. We also find temporal changes
of the association patterns for some variables. For instance, Airbnb has become
popular among areas with higher fraction of young residents.

\begin{table}[t]
\caption{Negative binomial regressions number of private/shared room hosts in a
census tract over time.}
\label{tab:nhost-other-year}
\begin{minipage}{\columnwidth}
\begin{center}
\begin{tabular}{l c c c c}
\toprule
& \multicolumn{4}{c}{\emph{Dependent variable: Number of hosts (private/shared room)}} \\
\cline{2-5}
                 & Before 2013                 & 2013                        & 2014                       & 2015 \\
\midrule
$popu$           & $\mwd0.223^{***}$   (0.012) & $\mwd0.194^{***}$ (0.012)   & $\mwd0.255^{***}$  (0.009) & $\mwd0.283^{***}$ (0.007) \\ 
$hotel$          & $\mwd0.028^{**\os}$ (0.009) & $\mwd0.034^{***}$ (0.009)   & $\mwd0.041^{***}$  (0.007) & $\mwd0.045^{***}$ (0.006) \\ 
$attraction$     & $\mwd0.063^{***}$   (0.008) & $\mwd0.052^{***}$ (0.008)   & $\mwd0.054^{***}$  (0.007) & $\mwd0.041^{***}$ (0.006) \\ 
$young$          & $   -0.006^{\ts}$   (0.016) & $\mwd0.006^{\ts}$ (0.016)   & $\mwd0.027^{*\ds}$ (0.013) & $\mwd0.058^{***}$ (0.011) \\ 
$nonwhite$       & $\mwd0.046^{*\ds}$  (0.018) & $\mwd0.008^{\ts}$ (0.019)   & $   -0.005^{\ts}$  (0.016) & $\mwd0.057^{***}$ (0.012) \\ 
$foreign$        & $   -0.039^{*\ds}$  (0.017) & $   -0.018^{\ts}$ (0.017)   & $   -0.007^{\ts}$  (0.014) & $\mwd0.028^{**\os}$ (0.011) \\ 
$owned$          & $   -0.094^{***}$   (0.024) & $   -0.080^{**\os}$ (0.025) & $   -0.089^{***}$  (0.020) & $\mwd0.007^{\ts}$ (0.016) \\ 
$housing\_value$ & $\mwd0.422^{***}$   (0.015) & $\mwd0.338^{***}$ (0.015)   & $\mwd0.281^{***}$  (0.013) & $\mwd0.205^{***}$ (0.011) \\ 
$rent$           & $\mwd0.131^{***}$   (0.020) & $\mwd0.173^{***}$ (0.020)   & $\mwd0.169^{***}$  (0.017) & $\mwd0.228^{***}$ (0.013) \\ 
$income$         & $   -0.579^{***}$   (0.028) & $   -0.546^{***}$ (0.029)   & $   -0.491^{***}$  (0.024) & $   -0.497^{***}$ (0.019) \\ 
$edu$            & $\mwd0.790^{***}$   (0.020) & $\mwd0.763^{***}$ (0.021)   & $\mwd0.732^{***}$  (0.017) & $\mwd0.685^{***}$ (0.014) \\ 
$employment$     & $\mwd0.032^{\ts}$   (0.020) & $\mwd0.049^{*\ds}$ (0.021)  & $\mwd0.069^{***}$  (0.017) & $\mwd0.082^{***}$ (0.013) \\ 
$arts$           & $\mwd0.266^{***}$   (0.012) & $\mwd0.269^{***}$ (0.013)   & $\mwd0.252^{***}$  (0.010) & $\mwd0.249^{***}$ (0.008) \\ 
Constant         & $   -2.604^{***}$   (0.016) & $   -2.692^{***}$ (0.017)   & $   -2.180^{***}$  (0.013) & $   -1.616^{***}$ (0.010) \\
\midrule
Observations     & $69,824$ & $69,824$ & $69,824$ & $69,824$ \\
Log Likelihood   & $-22,904.650$ & $-21,241.810$ & $-29,673.960$ & $-41,908.290$ \\
$\theta$         & $0.607^{***}$ (0.023) & $0.660^{***}$ (0.029) & $0.786^{***}$ (0.027) & $1.030^{***}$ (0.029) \\
AIC              & $45,837.300$ & $42,511.630$ & $59,375.920$ & $83,844.570$ \\
\bottomrule
\end{tabular}
\end{center}
\bigskip
\footnotesize\emph{Note:} \hfill $^{*}\text{p}<0.05$; $^{**}\text{p}<0.01$; $^{***}\text{p}<0.001$
\end{minipage}
\end{table}

\subsection{RQ3} \label{subsec:rq3}

Table~\ref{tab:new-review} provides the results from OLS regressions where the
dependent variables are monthly new reviews received by listings and the
independent variables are characteristics about the listing, its host, and the
CT where it is located. We include listing- and host-level features because the
number of new reviews is likely affected by, in addition to geolocation,
factors such as the number of amenities in the room, the length of membership
of the host, etc. We therefore need to control for these effects.

Regarding listing-level effects, listings tend to gain more new reviews if they
already have a larger number of existing reviews, a higher rating, or an
instant-book feature that allows guests to book the room directly without
host's approval. Regarding host-level variables, listings with a higher
response rate or faster response time tend to receive more new reviews.
However, after controlling for these effects, we find that entire home listings
in CTs with a higher median household income tend to accumulate more new
reviews, suggesting an advantage for SES-advantaged areas. Such association is
negative but not statistically significant for private/shared rooms.

\setlength{\tabcolsep}{16pt}
\begin{table}[t]
\caption{OLS regressions of number of monthly new reviews for entire home and
private room listings.}
\label{tab:new-review}
\begin{minipage}{\columnwidth}
\begin{center}
\begin{tabular}{l c c}
\toprule
& \multicolumn{2}{c}{\emph{Dependent variable: Number of monthly new reviews}} \\
\cline{2-3}
& Entire home & Private room \\
\midrule
$existing\_review$ & $\mwd0.905^{***}$   (0.007) & $\mwd1.046^{***}$ (0.013) \\
$rating$           & $\mwd0.378^{***}$   (0.008) & $\mwd0.498^{***}$ (0.014) \\
$amenities$        & $   -0.040^{***}$   (0.007) & $\mwd0.068^{***}$ (0.013) \\
$instant\_book$    & $\mwd0.158^{***}$   (0.007) & $\mwd0.208^{***}$ (0.012) \\
$photos$           & $   -0.034^{***}$   (0.007) & $\mwd0.006^{\ts}$ (0.012) \\
$host\_age$        & $   -0.354^{***}$   (0.007) & $   -0.478^{***}$ (0.013) \\
$host\_super$      & $\mwd0.167^{***}$   (0.007) & $\mwd0.071^{***}$ (0.012) \\
$host\_desp$       & $\mwd0.090^{***}$   (0.007) & $\mwd0.073^{***}$ (0.013) \\
$host\_resp\_rate$ & $\mwd0.039^{***}$   (0.007) & $\mwd0.068^{***}$ (0.014) \\
$host\_resp\_time$ & $\mwd0.270^{***}$   (0.008) & $\mwd0.304^{***}$ (0.014) \\
$popu$             & $   -0.004^{\ts}$   (0.007) & $   -0.035^{**\os}$ (0.012) \\
$young$            & $\mwd0.094^{***}$   (0.011) & $\mwd0.068^{***}$ (0.020) \\
$nonwhite$         & $   -0.012^{\ts}$   (0.008) & $   -0.050^{**\os}$ (0.015) \\
$foreign$          & $   -0.074^{***}$   (0.009) & $   -0.247^{***}$ (0.016) \\
$income$           & $\mwd0.070^{***}$   (0.015) & $   -0.036^{\ts}$ (0.028) \\
$edu$              & $   -0.052^{***}$   (0.011) & $   -0.129^{***}$ (0.020) \\
$employment$       & $\mwd0.020^{**\os}$ (0.008) & $\mwd0.033^{*\ds}$ (0.015) \\
$arts$             & $   -0.064^{***}$   (0.007) & $\mwd0.087^{***}$ (0.013) \\
$owned$            & $   -0.142^{***}$   (0.013) & $   -0.082^{***}$ (0.024) \\
$housing\_value$   & $\mwd0.043^{***}$   (0.011) & $\mwd0.262^{***}$ (0.020) \\
$rent$             & $   -0.048^{***}$   (0.012) & $   -0.083^{***}$ (0.021) \\
$hotel$            & $\mwd0.017^{*\ds}$  (0.007) & $\mwd0.070^{***}$ (0.012) \\
$attraction$       & $\mwd0.034^{***}$   (0.008) & $\mwd0.026^{\ts}$ (0.013) \\
Constant           & $\mwd1.701^{***}$   (0.007) & $\mwd2.077^{***}$ (0.012) \\
\midrule
Observations       & $75,485$ & $37,502$ \\
R$^{2}$            & $0.370$  & $0.325$ \\
Adjusted R$^{2}$   & $0.370$ & $0.324$ \\
Residual Std. Error & $1.757$ (df = 75461) & $2.257$ (df = 37478) \\
F Statistic        & $1,927.071^{***}$ (df = 23; 75461) & $783.752^{***}$ (df = 23; 37478) \\
\bottomrule
\end{tabular}
\end{center}
\bigskip
\footnotesize\emph{Note:} \hfill $^{*}\text{p}<0.05$; $^{**}\text{p}<0.01$; $^{***}\text{p}<0.001$
\end{minipage}
\end{table}

\section{Concluding Discussion}

In this paper, we study how social-economic statuses affect the participation
of service providers in the sharing economy and who actually benefit from it.
We focused on Airbnb---a pioneer yet less studied example of sharing economy
platforms---in the US as a case study. We used a large number of listings
located in the US and linked them with US Census data that contain demographic,
SES, and housing characteristics of census tracts.

We found that income has the most significant effect on residents'
participation in Airbnb; CTs with lower median household income are likely to
have more residents becoming Airbnb hosts. This association is persistent
across different forms of sharing (i.e., entire home or private/shared room)
and across years.

However, when examining the performance of hosts---measured by number of new
reviews in a month---we found that entire-home listings that are located in
areas with higher income have received more reviews, even after controlling for
listing- and host-level characteristics. These findings empirically demonstrate
that the advantage of SES-advantaged areas and the disadvantage of
SES-disadvantaged areas are present in the sharing economy.

Our work contributes to the literature on the understanding of supply side
actors of online markets. First, we provide the first extensive large-scale
quantitative analysis of the supply side of online markets from the SES
perspective. This is crucial to a comprehensive understanding of how these
markets function, especially given that existing studies either focus
predominantly on demographics, or are mainly qualitative, or of a small-scale.
Second, in line with previous interview-based studies which argue the incentive
role of monetary compensations, we demonstrate empirically that Airbnb indeed
appeal particularly to people in lower-income areas. Third, our results also
uncover that hosts in lower-income areas may not benefit from the market as
much as those in higher-income areas.

More broadly, our work also contributes to the studies examining the broader
societal impacts of online markets and digital platforms in general. Be these
impacts positive~\cite{Chen-value-2017} or negative~\cite{Chan-internet-2014,
Chan-crime-2016}, on established industries~\cite{Cramer-disruptive-2016,
Guttentag-disruptive-2015} or on individuals~\cite{Chan-internet-2014}, it is
incumbent upon researchers to continue this line of work to gradually open
these black-boxes and to provide empirical evidences that can inform policy.
Our work highlights a positive effect that Airbnb has attracted more people
from lower income areas.

On a related note, despite some critiques around sharing economy platforms like
Airbnb, they have various potentials to benefit the society. From the
customer's perspective, they make accommodations potentially more affordable to
tourists and have positive impact on local economic activities through
touristic related activities such as shopping, dining, and sightseeing. From
the service provider's perspective, they provide a way to allow people to
generate additional revenues. This may be more important for people from
lower-SES, corroborated by our results showing that Airbnb has attracted more
hosts in lower income areas.

We have focused on Airbnb in the US as a case study to examine how SES factors
impact the participation of service providers in the sharing economy and who
actually benefit from it. It remains to be seen whether our results can be
generalized to other countries or other marketplaces. As many online
marketplaces are alternatives to traditional markets but with a relatively low
entry barrier for service providers, we hypothesize that the potential monetary
gain will make them universally attractive to the lower-income community. Their
distinctions, however, may introduce some delicate differences. For example,
from a service provider's point of view, Airbnb and Uber may require different
levels of commitment of time, energy, or personal space, making one of them
more costly than the other, affecting participation.

Our work has clear policy implications. First, our results provide empirical
evidence that sharing economy platforms such as Airbnb may be particularly
attractive to low-SES individuals and provide additional revenue streams for
them. Although there have been studies showing the disruptive impact on
established industries~\cite{Cramer-disruptive-2016,
Guttentag-disruptive-2015}, policy makers who are currently debating whether
and how to accommodate these online platforms may find our results valuable
when considering their potential benefits. Second, that the advantage of
SES-advantaged areas might make hosts in higher-income areas benefit more calls
for a closer look before making any decisions on whom to tax and how much to
tax.

Our work also speaks to platform design. Platform designer may provide online
trainings for how to better use of the platform and how to deal with various
situations associated with managing listings, and more aim the platform at
those hosts with lower income. While having an instant-book feature helps
listings attract new reviews, enabling that for all listings would require much
work from the platform's side to ease any safety concerns hosts may have.

Our work is subject to the following limitations. First, despite that the
sample of listings analyzed here is the largest ever in the current literature,
we cannot ascertain how much it has covered all the active listings in the US
market, nor do we know, more importantly, if there is systemic bias in the
sample. This issue is raised because the data collection process involved
repeated queries of web pages displaying search results of listings in a
region, which were returned by Airbnb's search engine, which is a black-box.
Examining potential bias of this opaque, algorithmic system is an interesting
line of future work and also contributes to the increasingly important field of
algorithm auditing that aims to understand the impact of those systems on their
end-users (e.g.~\cite{Chen-peeking-2015}).

Second, the listings that are available to us are the ones that were active
during the data collection process. This means that, when examining how
association patterns change over time, we cannot observe listings that entered
and then exited the market before the time when data was collected, which may
to some extent dictate our results. Moreover, such limitation also means that
we cannot observe the group that either could not afford becoming a host or
choose not to. While this suggests that the lowest-income community might not
be better off, future work that incorporates decision to enter the market might
enhance our understanding.

Third, we have chosen to perform our analysis on the census tract level, given
the difficulty in accessing detailed demographic, SES, and housing data of each
individual host, although in theory we are interested in the association at the
individual level. If those data were available, more involved regression
designs that control for unobservable individual fixed effects might have been
used. While individual-level variables have larger variances than their
CT-level summaries, we expect our results to hold as regressions capture
average effects. Considering that individual-level data bring not only more
information but also noise, focusing on the census tract level may provide a
good balance.

\begin{acks}
We thank the anonymous referees for their helpful comments and suggestions
and Filippo Radicchi for excellent computing resources.
\end{acks}


\end{document}